\documentclass{bmcart}

\usepackage[utf8]{inputenc} 

\usepackage{multirow}
\usepackage{url}
\usepackage{graphicx,epstopdf}
\usepackage{latexsym}
\usepackage[tight]{subfigure}
\usepackage{algorithm}
\usepackage{algorithmic}
\usepackage{amsmath} 
\usepackage{amssymb}  
\usepackage{cases}
\usepackage{float}
\usepackage{array}

\def\includegraphics{}

\startlocaldefs
\endlocaldefs

\begin{document}

\begin{frontmatter}

\begin{fmbox}
\dochead{Research}


\title{Parallel Knowledge Embedding with MapReduce on a Multi-core Processor}


\author[
   addressref={aff1,aff2,aff3},                   
   corref={aff1},                              
   email={fanmiao.cslt.thu@gmail.com}  
]{\inits{MF}\fnm{Miao} \snm{Fan}}
\author[
   addressref={aff1,aff2,aff3},                   
]{\inits{QZ}\fnm{Qiang} \snm{Zhou}}
\author[
   addressref={aff1,aff2,aff3},                   
]{\inits{TFZ}\fnm{Thomas Fang} \snm{Zheng}}
\author[
      addressref={aff4,aff5,aff6},                  
]{\inits{RG}\fnm{Ralph} \snm{Grishman}}

\address[id=aff1]{
  \orgname{Center for Speech and Language Technology, Research Institute of Information Technology, Tsinghua University}, 
  \street{ROOM 1-303, BLDG FIT},                     %
  \postcode{100084}                                  
  \city{Beijing},                                    
  \cny{China}                                        
}

\address[id=aff2]{
  \orgname{Center for Speech and Language Technologies, Division of Technical Innovation and Development, Tsinghua National Laboratory for Information Science and Technology}, 
  \street{ROOM 1-303, BLDG FIT},                     %
  \postcode{100084}                                  
  \city{Beijing},                                    
  \cny{China}                                        
}
\address[id=aff3]{
  \orgname{Department of Computer Science and Technology, Tsinghua University}, 
  \street{ROOM 1-303, BLDG FIT},                     %
  \postcode{100084}                                  
  \city{Beijing},                                    
  \cny{China}                                        
}

\address[id=aff4]{
  \orgname{Proteus Group}, 
  \street{715 Broadway, Room 703},                     %
  \postcode{10003}                                  
  \city{New York City},                                    
  \cny{U.S.A.}                                        
}
\address[id=aff5]{
  \orgname{Department of Computer Science}, 
  \street{715 Broadway, Room 703},                     %
  \postcode{10003}                                  
  \city{New York City},                                    
  \cny{U.S.A.}                                        
}
\address[id=aff6]{
  \orgname{Courant Institute of Mathematical Sciences}, 
  \street{715 Broadway, Room 703},                     %
  \postcode{10003}                                  
  \city{New York City},                                    
  \cny{U.S.A.}
}


\end{fmbox}


\begin{abstractbox}

\begin{abstract} 
This article firstly attempts to explore parallel algorithms of learning distributed representations for both entities and relations in large-scale knowledge repositories with {\it MapReduce} programming model on a multi-core processor.

We accelerate the training progress of a canonical knowledge embedding method, i.e. {\it translating embedding} ({\bf TransE}) model, by dividing a whole knowledge repository into several balanced subsets, and feeding each subset into an individual core where local embeddings can concurrently run updating during the {\it Map} phase.

However, it usually suffers from inconsistent low-dimensional vector representations of the same key, which are collected from different {\it Map} workers, and further leads to conflicts when conducting {\it Reduce} to merge the various vectors associated with the same key.

Therefore, we try several strategies to acquire the merged embeddings which may not only retain the performance of {\it entity inference}, {\it relation prediction}, and even {\it triplet classification} evaluated by the single-thread {\bf TransE} on several well-known knowledge bases such as Freebase and NELL, but also scale up the learning speed along with the number of cores within a processor.

So far, the empirical studies show that we could achieve comparable results as the single-thread {\bf TransE} performs by the {\it stochastic gradient descend} (SGD) algorithm, as well as increase the training speed multiple times via adapting the {\it batch gradient descend} (BGD) algorithm for {\it MapReduce} paradigm.
\end{abstract}




\end{abstractbox}
%

\end{frontmatter}



\section{Introduction}
The emerging trend of learning distributed representations for both entities and relationships using knowledge repositories \cite{Bordes2011} without extra text has drawn much attention from academia as well as industry. The learnt embeddings can not only perform significant improvements on several subtasks for knowledge population such as {\it entity inference}, {\it relation prediction} and {\it triplet identification}, but also further facilitate other AI applications, such as {\it Question Answering Systems} and {\it Automatic Summarization Systems}. Almost all the studies of knowledge embedding leverage existing repositories such as Wordnet\footnote{\url{https://wordnet.princeton.edu/}} \cite{Miller1995}, Freebase\footnote{\url{https://www.freebase.com/}} \cite{Bollacker2007,Bollacker2008} and NELL \footnote{\url{http://rtw.ml.cmu.edu/rtw/}} \cite{carlson-aaai}, which contain billions of triplets, and each of them is represented by $<head\_entity, relation, tail\_entity>$ abbreviated as $<h, r, t>$. Furthermore, each knowledge base is considered as a huge directed graph where nodes are entities, edges are relations and a triplet $<h, r, t>$ implies a structure that there is a relationship $r$ between the entities $h$ and $t$, pointing from the head entity $h$ to the tail entity $t$. This hierarchical structure inspired Bordes et al. \cite{Bordes2013a} to interpret relationships as translation vectors operating between the embeddings of two entities. Hence, they have proposed a canonical model named {\bf TransE} (translating embedding) and led the prosperous research of knowledge embedding \cite{Fan-EtAl:2014:PACLIC,fan2015probabilistic,fan2015learning,DBLP:journals/corr/FanCHG15,DBLP:journals/corr/FanZZG15,D14-1167,DBLP:conf/aaai/WangZFC14} that focuses on devising models easy to train, containing reduced number of parameters and expected to scale up to very large knowledge bases.

However, to the best of our knowledge, the classical approach {\bf TransE} \cite{Bordes2013a} and its successors \cite{Fan-EtAl:2014:PACLIC,fan2015probabilistic,fan2015learning,DBLP:journals/corr/FanCHG15,DBLP:journals/corr/FanZZG15,D14-1167,DBLP:conf/aaai/WangZFC14} conventionally reported the outstanding results of experiments conducted on relatively small datasets such as {\bf WN100K} and {\bf FB150K}\footnote{We change the original names, {\bf WN} and {\bf FB15K}, of the two datasets to emphasize the sizes of them. }, but few of them mentioned the efficiency of training with really large scale datasets. The reason we believe is that all the models including the state-of-the-art, only demonstrated the training algorithms within the programming paradigm of single thread, which extremely seals the potential power of multi-core computing.

Multi-core processors enable the computation conducted simultaneously, but reform the paradigm of programming. An intuitive way of parallel computing is to divide the dataset and to feed each core with each subset to perform individual calculations. Therefore, the key challenges  of parallel computing with a multi-core processor are: how to schedule the computation tasks, maintain the algorithm of each core and synchronize the shared memory if necessary. To tackle these issues, Google researchers designed a well-known programming model called {\it MapReduce} \cite{dean2008mapreduce} for distributed computing, and it has been widely adopted by both commercial and open-source clusters. {\it MapReduce} generally provides two interfaces to programmers for adapting their single-thread algorithms:
\begin{itemize}
  \item {\it Map} is responsible for processing the input data, and emits a set of {\it intermediate} key/value pairs based on the algorithm devised by programmers.
  \item {\it Reduce} accepts an intermediate key and a set of values for the key in the meanwhile. It uses customized function to merge the values together, and finally produces zero or one output for that key.
\end{itemize}
During the period of operations, an independent master schedules multiple {\it Map} and {\it Reduce} workers to run concurrently and makes parallel computing possible.

So far, several quintessential machine learning models have already been adapted to the framework of {\it MapReduce} by Chu et al. \cite{chu2007map}. However, embedding-based models such as {\it word embedding} \cite{mikolov2013efficient} and {\it knowledge embedding} \cite{Bordes2011} are different from those traditional learning approaches like {\it logistic regression}, as the data itself constructs the parameter space of embedding-based models. For example, {\it knowledge embedding} models acquire the low-dimensional vector representation of each entity and relation within a repository, and if we split that knowledge base and distribute subsets into multiple {\it Map} workers, the parameter space is divided at the same time as well. It usually leads to inconsistent distributed representations of the same key when conducting {\it Reduce} with those intermediate key/value pairs emitted by many {\it Map} workers. The problem is still challenging and has not been properly solved yet.

Therefore, this paper firstly attempts to explore parallel algorithms of learning distributed representations for both entities and relations in large-scale knowledge repositories with {\it MapReduce} programming model on a multi-core processor.
We accelerate the training progress of the canonical knowledge embedding method, i.e. {\it translating embedding} ({\bf TransE}) model \cite{Bordes2013a}, by dividing a whole knowledge repository into several balanced subsets, and feeding each subset into an individual core where local embeddings can concurrently run updating during the {\it Map} phase.

To deal with the merging problem of {\it Reduce} phase caused by inconsistent low-dimensional vector representations of the same key collected from different {\it Map} cores,
we contribute several strategies to acquire the embeddings which may not only retain the performance of {\it entity inference}, {\it relation prediction}, and even {\it triplet classification} evaluated by the single-thread {\bf TransE} on several well-known knowledge bases such as Freebase and NELL, but also scale up the learning speed along with the number of cores within a processor.
So far, the empirical studies show that we could achieve comparable results as the single-thread {\bf TransE} performs by the {\it stochastic gradient descend} (SGD) algorithm as well as increase the training speed multiple times, via adapting the {\it batch gradient descend} (BGD) algorithm for {\it MapReduce} paradigm.

Here we arrange the subsequent content of this article as follows: Section 2 helps us to recap the {\bf TransE} model and its single-thread training algorithm based on {\it stochastic gradient descend} (SGD). Then, we adapt the single-thread algorithm into the framework of {\it MapReduce} in Section 3. Specifically, Section 3.1 talks about directly parallelizing the SGD-based algorithm within the {\it MapReduce} paradigm, and proposes three strategies of reducing the conflicted embeddings. Section 3.2, from another perspective, focuses on concurrently updating the gradients rather than the inconsistent embeddings to avoid conflicts, and suggests adapting the {\it batch gradient descend} (BGD) algorithm instead into the {\it MapReduce} paradigm. (TO BE CONTINUED...)

\section{Singlethread TransE}
Starting from 2011, when Bordes et al. \cite{Bordes2011} engaged in the representation learning of entities and relationships within knowledge bases, the literatures on knowledge embedding had been increasing, but few of them could support large-scale training within acceptable time due to the high complexity of their models. Until the year of 2013, Bordes et al. \cite{Bordes2013a} devised {\bf TransE}, a canonical model which pursued only one low-dimensional vector representation for each entity and each relationship. The reduced parameters required made it possible to scale up to very large datasets. Moreover, {\bf TransE} learnt those distributed representations with a single-layer energy-based model which was rather easier to train.

\subsection{Translating Embedding Model}
Inspired by the hierarchical structures of the relationship between two entities shown by Figure 1(a), Bordes et al. interpreted relationships as translations operating on entities and assumed that if a triplet $<h, r, t>$ holds, then ${\bf h} + {\bf r} \approx {\bf t}$. The boldface characters indicate $k$ dimensional vector representations ({\bf h}, {\bf r} and {\bf t}) taking values in $\mathbb{R}^k$. Equation (1) was derived from that assumption to measure the energy of a triplet $<h, r, t>$ as follows,
\begin{equation}
d({\bf h}, {\bf r}, {\bf t}) = ||{\bf h} + {\bf r} - {\bf t}||,
\end{equation}
in which we could take either $L_1$ or $L_2$-norm to measure the dissimilarity.
Given a training set $\Delta$ of triplets  $<h, r, t>$ composed of the set of entities $E$ ($h, t \in E$), and the set of relations $R$ ($r \in R$), we expected to discriminate the training triplets from corrupted triplets. One corrupted triplet $<h', r, t'>$ from its set $\Delta'$ was constructed according to Equation (2) by replacing either the head or tail entity randomly of a ground-truth triplet from the training set:
\begin{equation}
\Delta'_{(h, r, t)} = \{(h', r, t) | h' \in E ~\&~ h' \ne h \} \cup \{(h, r, t') | t' \in E ~\&~ t' \ne t\}.
\end{equation}
As we favored lower energy of training triplets than the corrupted ones, the objective was naturally to {\it minimize} the loss of a margin-based ranking function $\mathcal{L}$  over all the training triplets:
\begin{equation}
\mathcal{L} = \sum_{(h,r,t) \in \Delta} \sum_{(h',r,t')\in \Delta'_{(h,r,t)}} [\gamma + d({\bf h}, {\bf r}, {\bf t}) - d({\bf h'}, {\bf r}, {\bf t'})]_+,
\end{equation}
where $[x]_+$ was the hinge loss function that only considered the positive part of $x$ ($[x]_+ = \max\{0, x\}$), and $\gamma$ was that positive margin ($\gamma > 0$).

\subsection{SGD-based Singlethread Algorithm}
\begin{algorithm}
\caption{~~~~~The SGD-based Singlethread Learning Algorithm of {\bf TransE}}
\begin{algorithmic}[1]
\REQUIRE ~~\\
Training set $\Delta = \{(h, r, t)\}$, entity set $E$, relation set $R$;
dimension of embeddings $d$, margin $\gamma$, learning rate $\alpha$ for $\mathcal{L}$ , convergence threshold $\epsilon$, maximum epoches $n$.\\

\FOR{${\bf r} \in R$}
\STATE ${\bf r} := \text {Uniform} (\frac{-6}{\sqrt{d}}, \frac{6}{\sqrt{d}})$
\STATE ${\bf r} :=  \frac{{\bf r}}{|{\bf r}|} $\
\ENDFOR

\STATE $i := 0$
\WHILE{$Rel. loss > \epsilon$ and $i < n$ }
\FOR{${\bf e} \in E$}
\STATE ${\bf e} := \text {Uniform}(\frac{-6}{\sqrt{d}}, \frac{6}{\sqrt{d}})$
\STATE ${\bf e} :=  \frac{{\bf e}}{|{\bf e}|} $\
\ENDFOR

\FOR{$(h, r, t) \in \Delta$}
\STATE
\ENDFOR

\ENDWHILE
\\
\ENSURE ~~\\
All the embeddings of {\it e} and {\it r}, where ${\it e} \in E$ and ${\it r} \in R$.
\end{algorithmic}
\end{algorithm}
\section{MapReduce TransE}

\subsection{SGD-based MapReduce Paradigm}
\subsubsection{SGD-based Map phase}
\subsubsection{SGD-based Reduce phase}
\begin{itemize}
  \item {\it Random knowledge embedding}
  \item {\it Average knowledge embedding}
  \item {\it Mini-loss knowledge embedding}
\end{itemize}
\subsection{BGD-based MapReduce Paradigm}
\subsubsection{BGD-based Map phase}
\subsubsection{BGD-based Reduce phase}

\section{Experiment}



\section{Discussion}
\section{Conclusion}
This article firstly attempts to explore parallel algorithms of learning distributed representations for both entities and relations in large-scale knowledge repositories with {\it MapReduce} programming model on a multi-core processor.

We accelerate the training progress of a canonical knowledge embedding method, i.e. {\it translating embedding} ({\bf TransE}) model, by dividing a whole knowledge repository into several balanced subsets, and feeding each subset into an individual core where local embeddings can concurrently run updating during the {\it Map} phase.

However, it usually suffers from inconsistent low-dimensional vector representations of the same key, which are collected from different {\it Map} workers, and further leads to conflicts when conducting {\it Reduce} to merge the various vectors associated with the same key.

Therefore, we try several strategies to acquire the merged embeddings which may not only retain the performance of {\it entity inference}, {\it relation prediction}, and even {\it triplet classification} evaluated by the single-thread {\bf TransE} on several well-known knowledge bases such as Freebase and NELL, but also scale up the learning speed along with the number of cores within a processor.

So far, the empirical studies show that we could achieve comparable results as the single-thread {\bf TransE} performs by the {\it stochastic gradient descend} (SGD) algorithm, as well as increase the training speed multiple times via adapting the {\it batch gradient descend} (BGD) algorithm for {\it MapReduce} paradigm.

\begin{backmatter}

\section*{Acknowledgements}
This work is supported by China Scholarship Council, National Program on Key Basic Research Project (973 Program) under Grant 2013CB329304, National Science Foundation of China (NSFC) under Grant No.61433018 and No.61373075, Proteus Project of NYU. The first author conducted this research while he was a joint-supervision Ph.D. student of Tsinghua University and New York University.

\bibliographystyle{bmc-mathphys} 
\bibliography{bmc_article}      





\end{backmatter}
\end{document}